\documentclass[10pt, conference]{IEEEtran}

\usepackage{amssymb}
\usepackage{amsmath}

\PassOptionsToPackage{hyphens}{url}

\usepackage[colorlinks = true,
  linkcolor = blue,
  citecolor = blue,
  urlcolor = blue]{hyperref}

\usepackage{graphicx}
\usepackage{booktabs}
\usepackage{tabularx}
\usepackage{enumitem}
\usepackage{balance}

\begin{document}

\title{Undefined Behavior in C and C++: \\ An Experiment With Desktop Use Cases}

\author{
\IEEEauthorblockN{Jukka Ruohonen}
\IEEEauthorblockA{University of Southern Denmark \\
Email: juk@mmmi.sdu.dk}
\and
\IEEEauthorblockN{Krzysztof Sierszecki}
\IEEEauthorblockA{University of Southern Denmark \\
Email: krzys@mmmi.sdu.dk}
}

\maketitle

\begin{abstract}
Undefined behavior is idiomatic to C and C++ programming; such behavior is a use
of an erroneous program construct for which the languages impose no
requirements, such as integer overflows. The paper presents an empirical
experiment seeking to probe the extent of undefined behavior executing
underneath typical desktop use of a Linux distribution. The analysis is based on
an undefined behavior sanitizer implemented in a compiler. According to the
results, undefined behavior is common. By completing $59$ simple experimental
tasks, nearly $11$ thousand unique undefined behavior warnings were generated by
$32$ unique programs and libraries written in C or C++. Of these warnings, most
were associated with the Mesa graphics library and generated by interacting with
graphical user interfaces. Merely logging into the GNOME desktop environment
generated over $500$ unique warnings. Of all warnings, the clear majority was
about virtual table pointers. The associated stack traces were also lengthy in
general. With these and other results, the paper contributes to the empirical
literature on C and C++.
\end{abstract}


\section{Introduction}

Code sanitizers are auxiliary tools used by compilers for inserting
instrumentation code at runtime to detect bugs and other suspicious
behavior. Especially when used with other software testing techniques, including
fuzzing, sanitizers have been used to find countless vulnerabilities and
conventional, non-security bugs, including in operating system
kernels~\text{\cite{Ruohonen19RSDA, Song19}}. Also the undefined behavior
sanitizer (UBSan) has received appraisals from developers. Some teams ``have
found it to be invaluable to improve security''~\cite{Oracle21}. Although false
positives are a problem with all sanitizers, including ``harmless undefined
behavior instances'' in the present context, even simple exploratory exercises
with UBSan have caught many real bugs~\cite{Packard25}. Thus, UBSan has filled a
demand because traditionally---and possibly still so---undefined behavior was
difficult to identify and diagnose, causing a risk of developers incorrectly
brushing it off as ``GCC bugs''~\cite[p.~1]{Wang12}.

With these motivating introductory points in mind, the paper presents an
empirical experiment of undefined behavior in typical desktop use of open source
software (OSS) written in C and C++. The motivating question is: how much
undefined behavior is a hypothetical user using a desktop Linux witnessing? As
soon discussed in Section~\ref{sec: background}, the research questions
specified in Section~\ref{sec: research questions} can be further motivated by
many relevant points. Among these is a tangential question about how much
potential security and other risks is she taking?

The experiment is conducted with the UBSan implementation of the GNU Compiler
Collection (GCC). It resembles closely a similar UBSan implementation in the the
Clang front-end for the LLVM compiler back-end~\cite{Clang26}. In fact, even the
GCC's documentation~\cite{GCC26a} references Clang's documentation on its
sanitizer for undefined behavior. The focus on GCC alone can be further
justified on the grounds that the experiment took relatively long time to
conduct and results are exploratory, shedding light on the extent of undefined
behavior executing underneath conventional desktop use of Linux.

After the two already mentioned sections on the background, including related
work, and the research questions, the experimental setup is elaborated in
Section~\ref{sec: experimental setup}. Results are presented in
Section~\ref{sec: results}. Sections~\ref{sec: conclusion}, \ref{sec:
  limitations}, and \ref{sec: closing remarks} present the conclusion reached, a
few limitations, and closing remarks.

\section{Background and Related Work}\label{sec: background}

The old C99 standard defines undefined behavior as a ``behavior, upon use of a
nonportable or erroneous program construct or of erroneous data, for which this
International Standard imposes no requirements''~\cite[p.~4]{WG14}. Thus,
undefined behavior is essentially something that has not been specified as a
requirement. The examples are many, ranging from programming mistakes to
nonportable operations. Responses to undefined behavior range ``from ignoring
the situation completely with unpredictable results'' to ``terminating a
translation or execution''~\text{\cite[p.~4]{WG14}}. From a perspective of
developers, undefined behavior is thus something that should be avoided. Some,
but not all, undefined behavior is known to be associated with security
vulnerabilities~\cite{WG12}. Reliability and portability are further reasons for
seeking to eliminate undefined behavior.

At the same time, compilers are exploiting undefined behavior for aggressive
optimizations, which may yield surprising results for developers~\cite{Wang12,
  Wang16}. Undefined behavior is thus both a bug and a
feature~\cite[p.~345]{Hathorn15}. Yet, even in case a given undefined behavior
being optimized was not risky as such, it may be that the optimizations
themselves ``lead to unpredictable defects and vulnerabilities'', and any
undefined behavior ``not yet addressed may become the new major candidate for
attackers to leverage for nefarious
purposes''~\text{\cite[p.~9]{WG12}}. Compiler optimizations have indeed been
observed to sometimes cause security bugs because compilers generally assume
that a source code is free of undefined behavior~\cite{Xu23}. The point is
strengthened by results indicating that the performance gains from optimizations
involving undefined behavior are only small~\cite{Popescu25}. Regardless, these
points signify a general trade-off involving not only developers and compilers
but also standardization of the C and C++ programming languages.

Even when a language standard imposes explicit requirements and thus rules out
undefined behavior, the so-called Hyrum's law should be recalled---it states
that when the amount of users grows over time, ``it does not matter what you
promise in the contract: all observable behaviors of your system will be
depended on by somebody''~\cite[p.~8]{Winters20}. Furthermore, standards evolve
too; introducing a new feature for C or C++ may open a door for new undefined
behavior.

In addition to these related works about undefined behavior in C and C++, there
exists a lot of previous research on code sanitizers. Sanitizers specific to
addresses, memory, and threads have all been actively examined~\cite{Nong20,
  Ruohonen19RSDA, Vyukovaa20}. Also UBSan has been examined
previously~\cite{Xu23, Zhang21}. In fact, there is research demonstrating bugs
in also in many code sanitizers themselves, including but not limited to
UBSan~\cite{Li24}. Finally, it is worth mentioning existing and related
empirical research about Linux distributions and packages distributed via
them~\cite{Rocca25, Ruohonen25PST}. However, to the best of the authors'
knowledge, no directly comparable work existing about studying undefined
behavior with exploratory methods, which are about ``studying objects in their
\textit{natural setting} and letting the findings emerge from the
observations''~\cite[p.~12; italics added]{Wohlin24}. The emphasis on natural
setting is there to stress that the empirical experiment conducted is not about
testing a sample of open source software with UBSan. Instead, it is about
studying undefined behavior in a natural setting of desktop Linux use. Watching
a video on YouTube with a web browser is a natural way to use a desktop for
numerous people. This simple example also demonstrates methodological novelty;
watching a video on YouTube with a web browser executes a lot of code that is
external to~the web browser~in~itself.

\section{Research Questions}\label{sec: research questions}

The following research questions (RQs) are examined:

\vspace{2pt}
\begin{enumerate}[label={RQ$_\arabic{enumi}$:}, leftmargin=1.3cm]
\itemsep 5pt
\item{How much undefined behavior is detected with the GCC's UBSan
  implementation when completing typical tasks on a desktop Linux distribution?}
\item{Are some particular programs and libraries (PaLs) often used on a Linux
  desktop particularly prone to generate undefined behavior warnings?}
\item{Is undefined behavior in C and C++ more prevalent with some specific task
  categories, such as system administration, web browsing, or office work?}
\item{How long are the stack traces of undefined behavior warnings generated by
  UBSan on average?}
\item{What kind of undefined behavior is detected by the GCC's UBSan given the
  experimental tasks?}
\end{enumerate}

\vspace{2pt} On the one hand, the ``how much'' wording used for RQ$_1$ further
underlines the exploratory nature of the experiment. In other words: prior to
conducting the experiment and writing the paper, neither of the authors could
deduce about a sensible hypothesis regarding the potential amount of undefined
behavior---even after reading numerous papers on the topic, some of which are
referenced in Section~\ref{sec: background}. On the other hand, the answer to
RQ$_1$ signifies also the value of exploratory research; now there is a sensible
hypothesis for further work.

\section{Experimental Setup}\label{sec: experimental setup}

\subsection{Environment}\label{subsec: environment}

The data collection in May 2026 process followed closely the one used in recent
research~\cite{Ruohonen25PST}. After having installed the Gentoo's minimal
installation medium for the \texttt{x86\_64} instruction set
architecture~\cite{Gentoo26a} to a Qemu virtual machine, a desktop-specific
userland~\cite{Gentoo26b}, or ``stage'' in the Gentoo's parlance, was installed
and a profile was subsequently specified to be
GNOME-specific.\footnote{~\texttt{default/linux/amd64/23.0/desktop/gnome
  (stable)}} During the installation process, all packages recommended for a
desktop system were installed according to a manual
provided~\cite{Gentoo26c}. The final step was the installation of additional
packages, including the GNOME desktop environment itself. Regarding the other
packages, these can be deduced from Table~\ref{tab: tasks} in
Subsection~\ref{subsec: experimental tasks}.

Unlike in related research~\cite{Ruohonen25PST}, Gentoo's default compiling
options were used. These include optimizations, which have also been recommended
for UBSan~\cite{Oracle21}. Given guidelines~\cite{Gentoo26d} and the Gentoo's
\texttt{make.conf} configuration file for building packages, the following
options were thus used for compiling:

\begin{small}
\begin{verbatim}
  COMMON_FLAGS="-march=native -O2 -pipe"
  CFLAGS="${CFLAGS} -fsanitize=undefined"
  CXXFLAGS="${CXXFLAGS} -fsanitize=undefined"
  LDFLAGS="${LDFLAGS} -fsanitize=undefined"
  USE="${USE} custom-cflags network-manager \
             -clang -luajittex"
\end{verbatim}
\end{small}

With these options, all packages in the installation were recompiled with the
GCC version 15.2.1 released in 2025.\footnote{~This GCC version uses the C23 and
C++17 language versions for C and C++ by default~\cite{GCC26b, GCC26c}, although
it may be that some packages specify their own flags also in this
regard. Furthermore, the sanitizer option had to be omitted for the
\texttt{app-arch/xz-util}, \texttt{app-text/texlive-core},
\texttt{dev-qt/qtbase}, \texttt{dev-qt/qtquick3d}, \texttt{kde-frameworks/kio},
\texttt{net-libs/webkit-gtk}, and \texttt{x11-libs/gtk+} packages because these
failed to build with it.} Another configuration used was about UBSan's runtime
behavior. Similarly to panic-causing assertions in
kernels~\cite{Ruohonen25ICTSSa}, and to ease debugging, some have preferred
UBSan's option to abort the execution of a process upon it triggering undefined
behavior~\cite{RedHat21}. As the intention of the experiment is empirical, such
options were not used. Instead, the following two configuration options were
embedded to a global environmental~variable:

\begin{small}
\begin{verbatim}
  UBSAN_OPTIONS="log_path=$HOME/ubsan/ubsan: \
                 print_stacktrace=1"
\end{verbatim}
\end{small}

By implication, the GCC's UBSan implementation logs at runtime each detected
undefined behavior into a file \texttt{\$HOME/ubsan/ubsan.[pid]}, where
\texttt{[pid]} denotes the process identifier number of the corresponding
process having caused the given undefined behavior. These files contain also
stack traces associated with the specific binaries having caused undefined
behavior logging. After parsing and quantification, the logs constitute the
empirical materials for the experiment.

\subsection{Experimental Tasks}\label{subsec: experimental tasks}

The experiment was conducted by completing the simple tasks from
$\textmd{T}_{s1}$ to $\textmd{T}_{c11}$ enumerated in Table~\ref{tab:
  tasks}. After completing a task, the log files outputted by UBSan, if any,
were moved to a specific directory used for parsing in order to avoid
mistakes. For the web browsing tasks, all caches and configuration files were
deleted between the tasks with the Epiphany browser.


\begin{table}[p!]
\centering
\caption{Experimental Tasks}
\label{tab: tasks}
\begin{tabular}{ll}
\toprule
$\textmd{T}_{ij}$ & Description \\
\hline
\\
& \underline{System administration:} \\
$\textmd{T}_{s1}$ & Boot the system and login to a terminal as \texttt{root} \\
$\textmd{T}_{s2}$ & Boot the system and login to GNOME as a normal user \\
$\textmd{T}_{s3}$ & Execute \texttt{htop} \\
$\textmd{T}_{s4}$ & Execute \texttt{joe \$HOME/.bashrc} \\
$\textmd{T}_{s5}$ & Execute \texttt{nano \$HOME/.bashrc} \\
$\textmd{T}_{s6}$ & Execute \texttt{emerge www-client/lynx} \\
$\textmd{T}_{s7}$ & Execute \texttt{ssh-keygen -t rsa -b 4096} \\
$\textmd{T}_{s8}$ & Execute \texttt{ssh-keygen -t ecdsa -b 521} \\
$\textmd{T}_{s9}$ & Execute \texttt{useradd foo \&\& userdel foo} \\
$\textmd{T}_{s10}$ & Execute \texttt{tar cvf archive.tar /usr/share/info} \\
$\textmd{T}_{s11}$ & Capture live traffic with Wireshark for fifteen seconds \\
$\textmd{T}_{s12}$ & Execute \texttt{openssl req -x509 -newkey rsa:4096} \\
$\textmd{T}_{s13}$ & Execute \texttt{tcpdump -c 5 \& ping google.com -c 5} \\
$\textmd{T}_{s14}$ & Open the GNOME's graphical printer configuration application \\
\hline
\\
& \underline{Desktop and multimedia:} \\
$\textmd{T}_{d1}$ & Open and close the classical xterm terminal emulator \\
$\textmd{T}_{d2}$ & Open and close the GNOME's default terminal (Console) \\
$\textmd{T}_{d3}$ & Open the Cheese application for web cameras \\
$\textmd{T}_{d4}$ & Open the Chess game shipped in GNOME \\
$\textmd{T}_{d5}$ & Play a game of Five or More shipped in GNOME \\
$\textmd{T}_{d6}$ & Play a game of Five-in-a-row shipped in GNOME \\
$\textmd{T}_{d7}$ & Open the weather widget application for GNOME \\
$\textmd{T}_{d8}$ & Open the GNOME's GNote note taking application \\
$\textmd{T}_{d9}$ & Open an image with the Eye of GNOME image viewer \\
$\textmd{T}_{d10}$ & Open and close the GNOME's application for contacts \\
$\textmd{T}_{d11}$ & Open Rhythmbox and listen a minute of Radio Paradise \\
$\textmd{T}_{d12}$ & Convert a PNG file with ImageMagick to the GIF format \\
$\textmd{T}_{d13}$ & Open the graphical system monitor application for GNOME \\
$\textmd{T}_{d14}$ & Use MPlayer to play a video encoded in the H.264 format \\
$\textmd{T}_{d15}$ & Open the clock application for GNOME start a stopwatch \\
$\textmd{T}_{d16}$ & Write ``hello world'' with the text editor for GNOME \\
$\textmd{T}_{d17}$ & Calculate $2 + 2$ with the calculator application for GNOME \\
$\textmd{T}_{d18}$ & Visit fifteen directories with the GNOME's Files application \\
$\textmd{T}_{d19}$ & Change the wallpaper with the GNOME's settings application \\
$\textmd{T}_{d20}$ & Open the graphical disk usage analyzer application for GNOME \\
\hline
\\
& \underline{Web browsing:} \\
$\textmd{T}_{w1}$ & Execute \texttt{curl http://example.com} \\
$\textmd{T}_{w2}$ & Execute \texttt{wget http://example.com} \\
$\textmd{T}_{w3}$ & Open Lynx and visit \texttt{https://ieee.org} \\
$\textmd{T}_{w4}$ & Open Epiphany and visit \texttt{https://ieee.org} \\
$\textmd{T}_{w5}$ & Open Epiphany and watch a random video on YouTube \\
\hline
\\
& \underline{Office work:} \\
$\textmd{T}_{o1}$ & Open this paper with Evince \\
$\textmd{T}_{o2}$ & Open this paper with Okular \\
$\textmd{T}_{o3}$ & Open this paper with MuPDF \\
$\textmd{T}_{o4}$ & Open the GnuCash accounting application \\
$\textmd{T}_{o5}$ & Open AbiWord and write ``undefined behavior'' \\
$\textmd{T}_{o6}$ & Open Gnumeric and write ``123'' to the cell A1 \\
$\textmd{T}_{o7}$ & Open the Mutt email client and draft an email \\
$\textmd{T}_{o8}$ & Open the Evolution email client and draft an email \\
$\textmd{T}_{o9}$ & Write ``new event'' to 12 May 2026 in the GNOME's calendar \\
\hline
\\ 
& \underline{Scientific work:} \\
$\textmd{T}_{c1}$ & Run \texttt{git init .} \\
$\textmd{T}_{c2}$ & Open the GTKWave application \\
$\textmd{T}_{c3}$ & Generate a simple plot with gnuplot \\
$\textmd{T}_{c4}$ & Generate a simple plot with Matplotlib \\
$\textmd{T}_{c5}$ & Calculate a matrix multiplication with Octave \\
$\textmd{T}_{c6}$ & Compile this paper with \texttt{pdflatex} \\
$\textmd{T}_{c7}$ & Transform the PDF of this paper with \texttt{pdf2ps} \\
$\textmd{T}_{c8}$ & Execute \texttt{pdfunite 1.pdf 2.pdf out.pdf} \\
$\textmd{T}_{c9}$ & Open the \LaTeX~source of this paper with XEmacs \\
$\textmd{T}_{c10}$ & Open the \LaTeX~source of this paper with Texmaker \\
$\textmd{T}_{c11}$ & Open R and run a command \texttt{plot(rnorm(1234))} \\
\bottomrule
\end{tabular}
\end{table}

The experiment can be classified as a technology-oriented quasi-experiment; the
objects are all software and no randomization was used for the tasks
experimenting with the objects~\cite[pp.~73--75]{Wohlin24}.\footnote{~The
authors are well-aware of the regrettable misuse of the term experiment in
software engineering research~\cite{Ayala22}. Therefore, to further clarify, the
experiment involves manipulation because the tasks directly influence the
results; hence, the technology-oriented quasi-experiment cannot be said to be an
observational study. There is also some limited comparison control.}  The
experiment also resembles black-box manual testing, the flexibility of which has
been perceived as an important benefit among practitioners whose experience and
time constraints have often guided test case selection~\cite{Haas21}. The point
about time constraints translates into a feasibility constraint in the present
context; it is not feasible to test a whole operating system running a desktop
environment. Because there is no directly comparable related work, nor is it
possible to rely on prior knowledge---a tactic often used for selecting and
prioritizing manual test cases~\text{\cite{Haas21, Hemmati17}}. Therefore, the
tasks were aligned toward the wording in the introduction about ``conventional
desktop use''. Watching YouTube videos is a conventional desktop use for
countless of people, whereas Matplotlib, Python, R, and \LaTeX~are used by many
scientists daily. The tasks were also designed with a focus on large PaLs,
which, by hypothesis, should be more prone to undefined behavior due to their
large code bases.

It can also be mentioned that all tasks were specified prior to having the
experimental Gentoo installation ready. Therefore, the experiment does not
resemble exploratory testing; cf.~\cite{Itkonen07}. Finally, regarding the
notion of black-box testing, it can be remarked that the authors have knowledge
about C, C++, and OSS, including its development, but no actual source code was
examined before, during, or after the experimental tasks.

\subsection{Parsing and Methods}\label{subsec: parsing}

The first three research questions, $\textmd{RQ}_1$, $\textmd{RQ}_2$, and
$\textmd{RQ}_3$, require only counting the amount of unique undefined behavior
warnings generated by UBSan. Uniqueness is ensured because the GCC's UBSan
implementation reports each warning in a PaL with references to a specific
source code file and a line number in it. Ensuring uniqueness is also important
because redundancy is a known problem with many sanitizers, including UBSan;
same checks are often repeatedly checked~\cite{Zhang21}. Such redundant checks
are not included in the results reported---each undefined behavior warning
observed is specific to a given C or C++ source code file and line in it.

Regarding $\textmd{RQ}_2$, the counting is done by recording the immediately
triggering PaLs based on the stack traces. Given parsing of lines containing
\texttt{\#0}, the following stripped snippet would thus yield two unique
warnings for the Gentoo's \texttt{app-editors/joe} package, which is identified
as \texttt{joe}:

\begin{small}
\begin{verbatim}
[...]

cclass.c:1032:16: runtime error: left shift \
      of negative value -32
  #0 [...] (/usr/bin/joe+0xc713) [...]
  #1 [...] (/usr/bin/joe+0x1fb1c3) [...]
  #2 [...] (/usr/bin/joe+0xcdb0) [...]
  #3 [...] (/usr/lib64/libc.so.6+0x273ea) [...]
  #4 [...] (/usr/lib64/libc.so.6+0x2749a) [...]
  #5 [...] (/usr/bin/joe+0x115e4) [...]

hash.c:22:10: runtime error: left shift [...] \
     cannot be represented in type 'long int'
  #0 [...]  (/usr/bin/joe+0x88be) [...]
  #1 [...]  (/usr/bin/joe+0x777d2) [...]

[...]
\end{verbatim}
\end{small}

Regarding $\textmd{RQ}_3$ and with a reference to the earlier snippet in
Subsection~\ref{subsec: environment} about \texttt{UBSAN\_OPTIONS}, the unique
warnings in all log files generated by UBSan are counted for each task. As for
$\textmd{RQ}_4$, the above stripped snippet suffices to exemplify the
parsing; the length of the first unique warning generated by executing the
\texttt{joe} editor is taken to be five.

\begin{table}[th!b]
\centering
\caption{Categorization of Warnings}
\label{tab: warning categories}
\begin{tabular}{ll}
\toprule
Category & Strings \\
\hline
Buffer overflows & $\bullet$~``\textit{with insufficient space for an object of type}'' \\
\cmidrule{1-1}
Integer overflows & $\bullet$~``\textit{signed integer overflow}'' \\
\cmidrule{1-1}
Shift operators & $\bullet$~``\textit{shift exponent}'' \\
& $\bullet$~``\textit{left shift of}'' \\
\cmidrule{1-1}
Misaligned accesses & $\bullet$~``\textit{store to misaligned address}'' \\
& $\bullet$~``\textit{load of misaligned address}'' \\
& $\bullet$~``\textit{member access within misaligned address}'' \\
\cmidrule{1-1}
NULL pointers & $\bullet$~``\textit{null pointer passed as argument}'' \\
& $\bullet$~``\textit{member access within null pointer of type}'' \\
\cmidrule{1-1}
Type mismatches & $\bullet$~``\textit{does not point to an object of type}'' \\
\cmidrule{1-1}
Variable length arrays & $\bullet$~``\textit{variable length array bound}'' \\
\cmidrule{1-1}
Virtual pointers & $\bullet$~``\textit{object has invalid vptr}'' \\
\bottomrule
\end{tabular}
\end{table}

Then, with respect to $\textmd{RQ}_5$, the warnings are categorized into eight
groups shown in Table~\ref{tab: warning categories}. Most of the categories have
been discussed in the literature~\cite{Wang12}. If a given string shown in the
second column of the table appears anywhere in a warning, the warning is
categorized into the corresponding category.


\section{Results}\label{sec: results}

Undefined behavior is common. Out of the $59$ experimental tasks conducted, as
many as $36$ generated at least one warning. In other words, UBSan outputted one
or more warnings for about $61\%$ of the simple experimental tasks in
Table~\ref{tab: tasks}.

\begin{figure}[th!b]
\centering
\includegraphics[width=\linewidth, height=4.5cm]{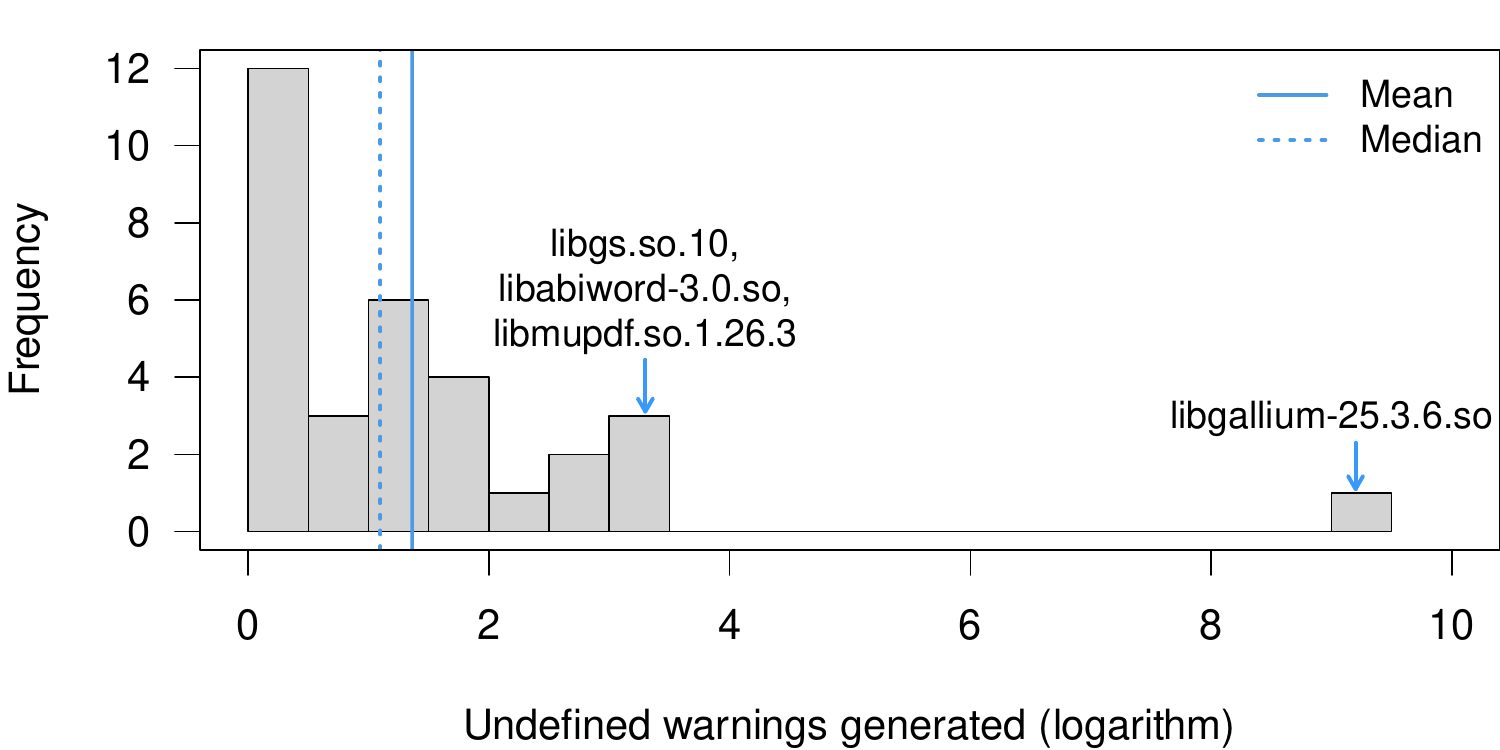}
\caption{Undefined Warnings Generated by PaLs}
\label{fig: warnings pals}
\end{figure}

\begin{figure}[th!b]
\centering
\includegraphics[width=\linewidth, height=6cm]{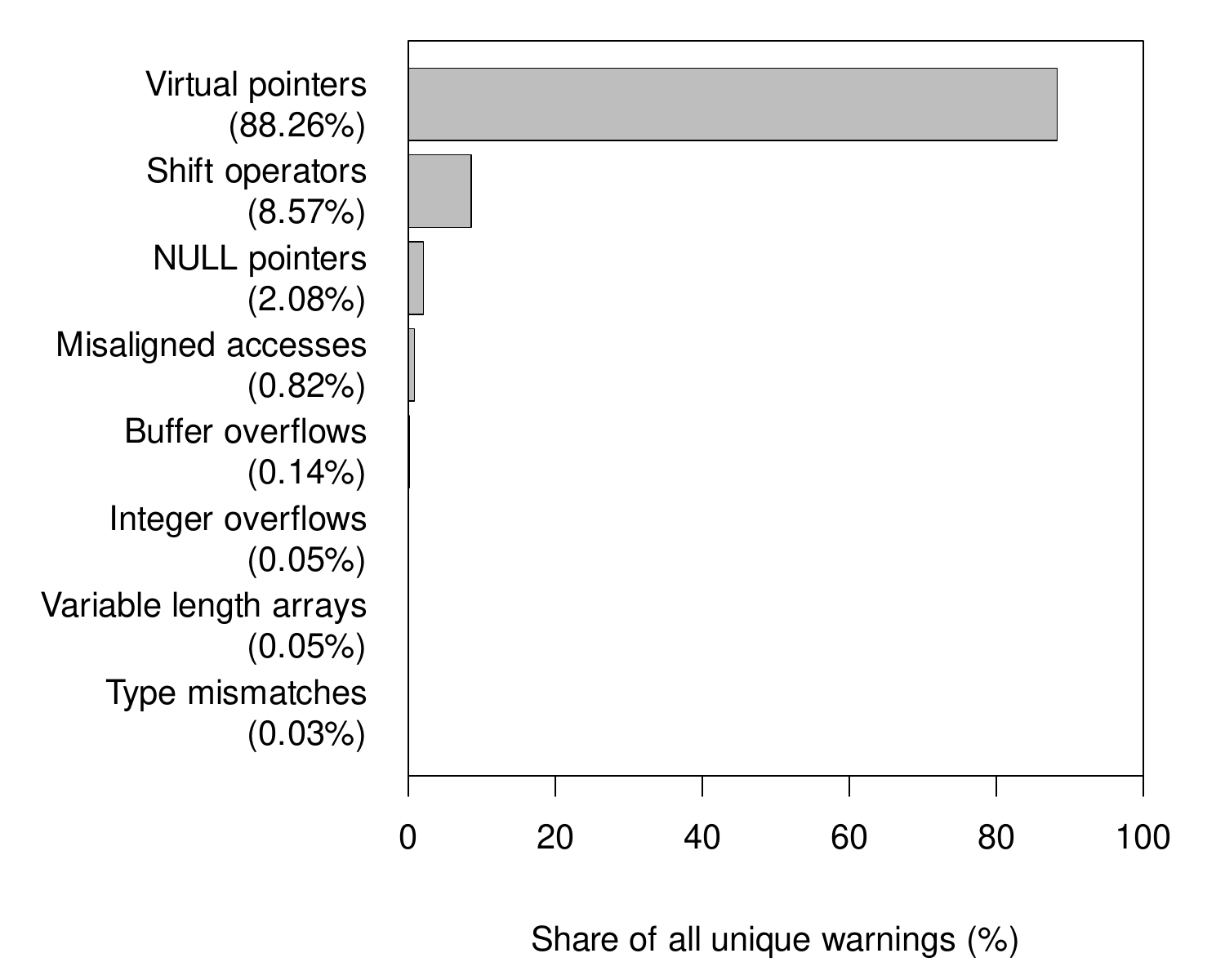}
\caption{Undefined Warning Categories}
\label{fig: warnings categories}
\end{figure}

In total, the experimental tasks generated $10,914$ unique undefined behavior
warnings. Yet, as can be deduced from Fig.~\ref{fig: warnings pals}, these
warnings are extremely unequally distributed across the $32$ unique PaLs that
generated undefined behavior warnings from the $59$ experimental tasks
conducted. In fact, \texttt{libgallium-25.3.6.so}, which is part of the
low-level Mesa graphics library, accounted for about $98\%$ of all unique
warnings. The Mesa's $10,723$ unique warnings are associated with $20$
experimental tasks. Even without saying anything about true and false positives,
these numbers underline that Mesa is frequently executed on a desktop Linux. At
the other end of the empirical distribution in Fig.~\ref{fig: warnings pals},
about 38\% of the unique PaLs identified generated just one unique warning.

Regarding the types of warnings outputted by UBSan, the overwhelming majority
are about virtual pointer tables used by compilers for C++. As seen from
Fig.~\ref{fig: warnings categories}, also bitwise shift operators were
relatively common. Two of the tasks, $\textmd{T}_{d14}$ and $\textmd{T}_{c7}$,
generated warnings hinting about potential buffer overflows. The warnings
associated with these two tasks were further associated with
\texttt{libavformat.so.60} and \texttt{libgs.so.10}, a multimedia library and a
library for Ghostscript. Otherwise, a notable takeaway from Fig.~\ref{fig:
  warnings categories} is that signed integer overflows were only seldom
detected by UBSan.

\begin{figure*}[th!b]
\centering
\includegraphics[width=\linewidth, height=5cm]{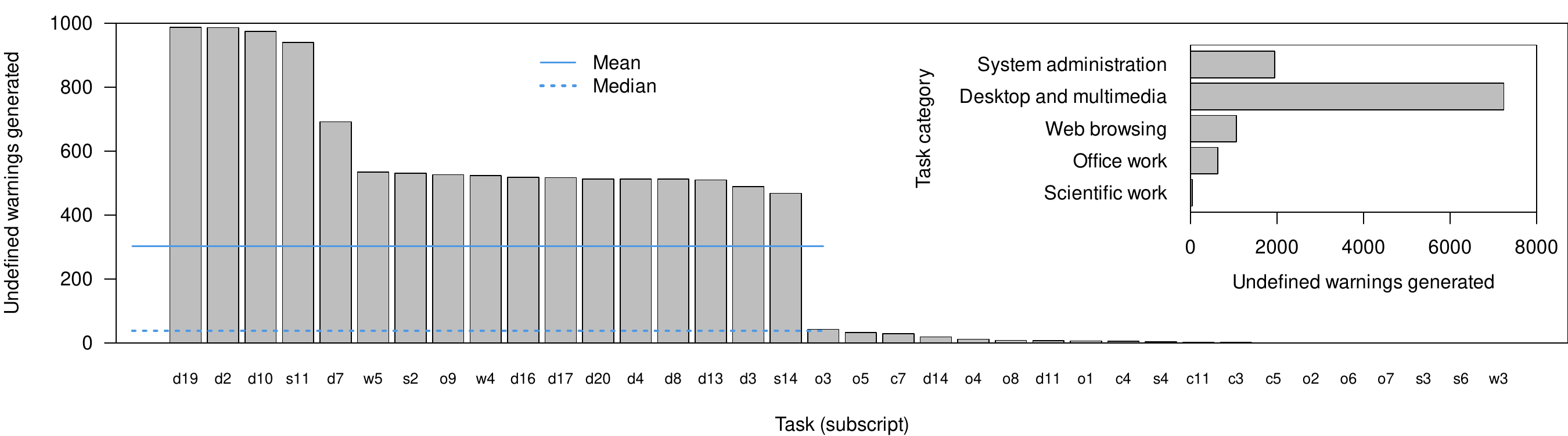}
\caption{Undefined Warnings Generated by Tasks (only tasks having generated at least one warning are shown)}
\label{fig: warnings tasks}
\end{figure*}

\begin{figure}[th!b]
\centering
\includegraphics[width=\linewidth, height=12.7cm]{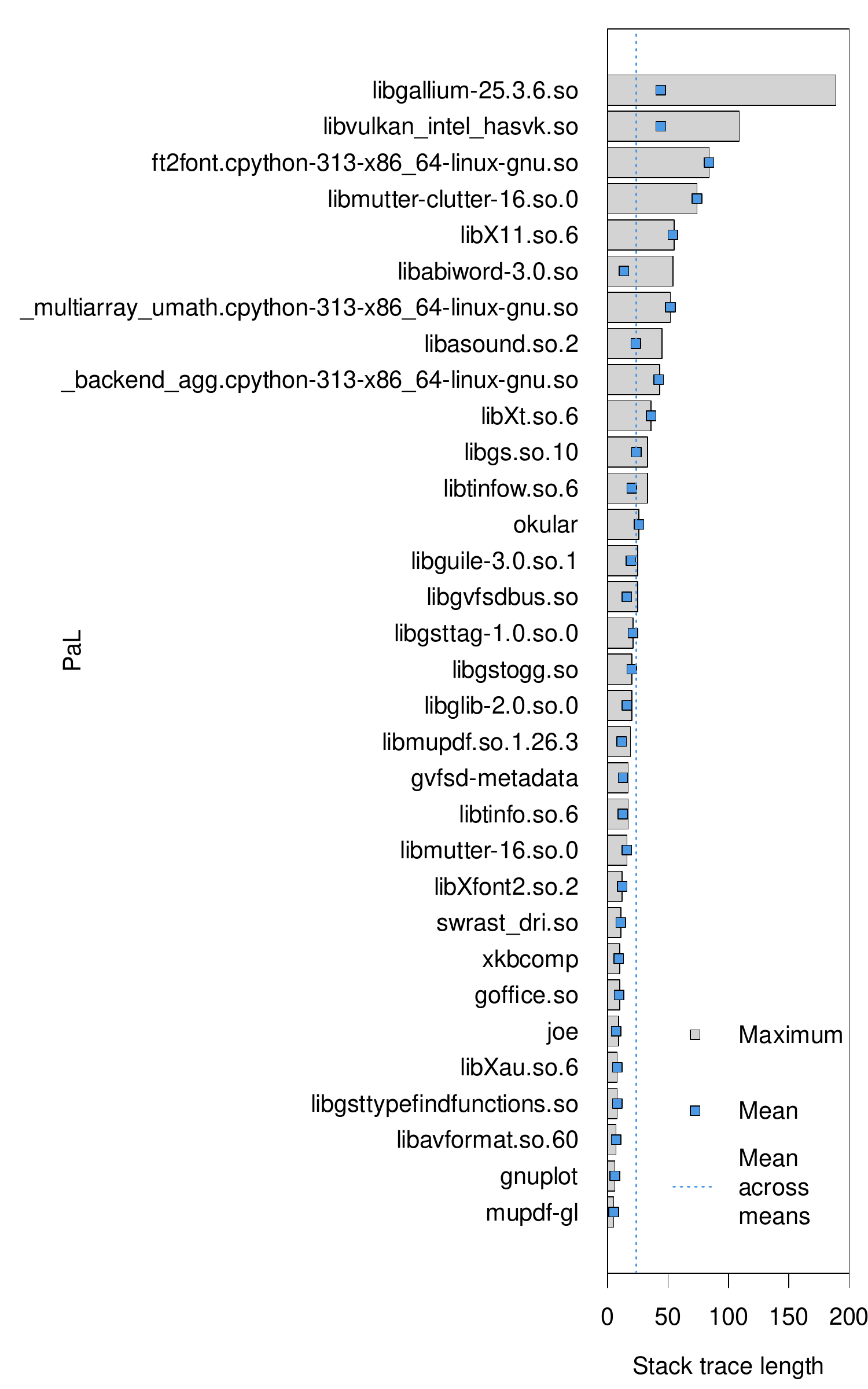}
\caption{Trace Lengths Across PaLs Having Generated One or More Warnings}
\label{fig: traces pals}
\end{figure}

\begin{figure}[th!b]
\centering
\includegraphics[width=\linewidth, height=12.7cm]{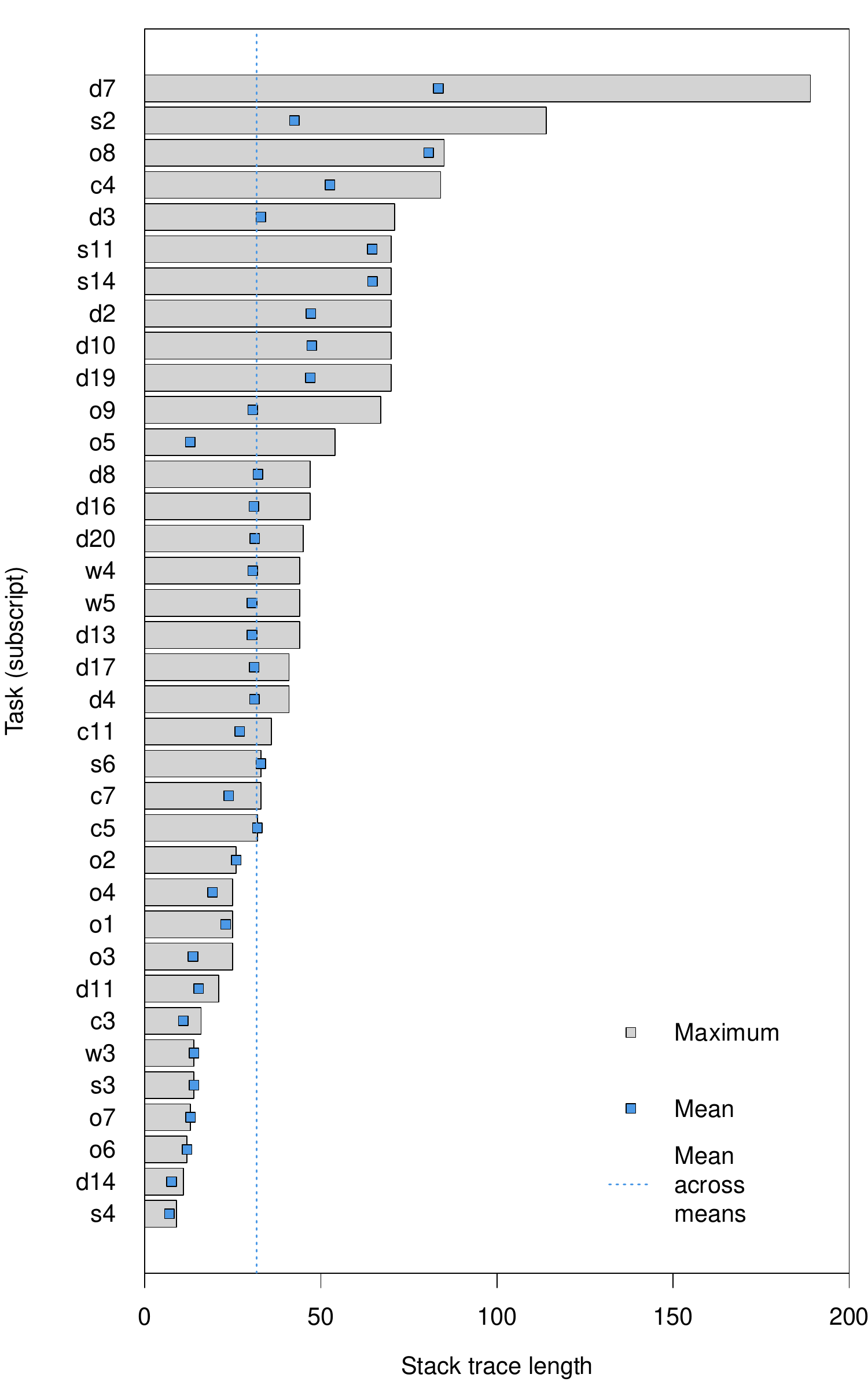}
\caption{Trace Lengths Across Tasks Having Generated One or More Warnings}
\label{fig: traces tasks}
\end{figure}

The distribution of the warnings is unequal also across the experimental
tasks. Although the tasks are not strictly comparable in statistical terms, most
(about 66\%) of the unique undefined behavior warnings were associated with the
desktop and multimedia tasks. As seen from the smaller inner plot in
Fig.~\ref{fig: warnings tasks}, also the system administration tasks generated a
lot of warnings. However, as can be deduced from the outer plot in the figure,
most of the warnings generated by the system administration tasks were related
to graphical user interfaces (in particular, $\textmd{T}_{s11}$,
$\textmd{T}_{s2}$, and $\textmd{T}_{s14}$). Merely logging in to the GNOME
desktop environment and trying to configure a printer generated almost one
thousand unique warnings. Also the many warnings generated by UBSan when using
the graphical user interface of Wireshark to record network traffic are worth
noting due to the program's history with
vulnerabilities.\footnote{~\url{https://www.cvedetails.com/version-list/0/8292/1/}}
The larger outer plot in Fig.~\ref{fig: warnings tasks} also reiterate the
earlier point about Mesa. In particular, the desktop use cases
$\textmd{T}_{d2}$, $\textmd{T}_{d10}$, and $\textmd{T}_{d19}$ all generated a
lot of unique undefined behavior warnings associated with
the noted \texttt{libgallium-25.3.6.so} library.

Mesa stands out also in terms of the stack trace lengths. When the maximum trace
lengths are considered, Mesa's two libraries take the topmost places in the
ranking shown in Fig.~\ref{fig: traces pals}. If mean lengths are considered,
the Mesa's Gallium library and \texttt{libvulkan\_intel\_hasvk.so} do not rank
as high, indicating a large variance in the traces and their lengths generated
by these two libraries. In contrast, the variance is small for the traces
generated by \texttt{ft2font.cpython-313-x86\_64-linux-gnu.so},
\texttt{libmutter\-clutter-16.so.0}, and \texttt{libX11.so.6}, among a few other
PaLs. In other words: for the three noted PaLs, the stack traces have been
consistently lengthy.

When the trace lengths are considered in terms of the experimental tasks, it is
interesting that a simple weather widget (as per the $\textmd{T}_{d7}$ task)
generated the lengthiest traces. When maximums are considered, the second place
is taken by $\textmd{T}_{s2}$, the task of logging into the GNOME desktop
environment. When compared to the weather widget, this observations seems more
logical because a lot of code is executed once a user logs into GNOME. Although
$\textmd{T}_{o8}$, the task involving the Evolution email client, takes the
third place in Fig.~\ref{fig: traces tasks}, the trace lengths do not seem to
consistently vary across program sizes. For instance, the tasks associated with
Matplotlib and the Cheese web camera application ($\textmd{T}_{c4}$ and
$\textmd{T}_{d3}$, respectively) also generated a few particularly lengthy stack
traces.

All in all, these observations can be used for supporting a tentative argument
that undefined behavior may not correlate \textit{as strongly} with software
size as the volume of real, verified bugs is known to correlate;
cf. \cite{Zhang09}, among others. A further point is that even simple graphical
user interfaces and small programs in general can generate a lot of undefined
behavior that is attributable to low-level libraries written in C and
C++. Therefore, it can also be argued that these are the areas to which effort
should be devoted for reducing the volume of undefined behavior executing
underneath typical desktop use of Linux. Debugging is required also for
verifying whether the warnings observed and generated by the GCC's UBSan are
actual true positives. That said, the point is easy to raise but likely
difficult in practice because undefined behavior is generally difficult to
learn, debug, and avoid even by experienced programmers~\cite{Xu23}. As with the
GCC's some other options~\cite{Ruohonen25PST}, another argument can therefore be
raised that improving the warnings outputted by UBSan might pay~off.

\section{Conclusion}\label{sec: conclusion}

The answers to the five research questions can be summarized as follows:
undefined behavior is very common with common desktop use cases
($\textmd{RQ}_1$), but some programs and libraries---particularly those
associated with Mesa---are more prone to generate undefined behavior warnings
($\textmd{RQ}_2$); undefined behavior is also more common with desktop and
multimedia tasks involving graphical user interfaces ($\textmd{RQ}_3$); the
stack traces associated with undefined behavior warnings are lengthy on average,
although also these vary across programs and libraries ($\textmd{RQ}_4$); and
virtual table pointers \text{and---to} a much lesser extent---shift operators,
NULL pointers, and misaligned accesses are behind most of the warnings
observed~($\textmd{RQ}_5$).

\section{Limitations}\label{sec: limitations}

The following four limitations should be acknowledged:

\vspace{2pt}
\begin{enumerate}
\itemsep 5pt
\item{Selection bias is a typical issue in software engineering
  quasi-experiments~\cite{Kampenes09}. Even though an attempt was made to
  slightly control it by allowing both authors to suggest tasks prior to
  conducting the experiment, it must be acknowledged that the tasks may not
  represent the tasks typical to desktop Linux use among people using desktop
  Linux. To address the limitation, a survey could be used in further work about
  typical desktop use cases. Even then, it should also be remarked that some
  packages failed to compile with UBSan---among them, notably, Firefox and
  LibreOffice. These had to be thus omitted.}
\item{UBSan is known to miss a lot of undefined behavior~\cite{Xu23}. Thus,
  particularly the answer to $\textmd{RQ}_1$ is only tentative. In~reverse:
  false positives are presumably also present.}
\item{Many of the technical limitations elaborated in closely related existing
  work~\cite{Ruohonen25PST} apply also the current work. For instance, some
  packages are known to force their own compiler flags, ignoring those specified
  in a distribution~\cite{Rocca25}. Such technical details may affect all of the
  RQs, although the effect can be argued to be only small. The bias was also
  mitigate by using the USE flag \texttt{custom-cflags} as a global option for
  all packages.}
\item{Many undefined behaviors are specific to hardware; an undefined behavior
  in some instruction set architecture may not be undefined behavior in some
  other architecture~\cite{Wang12, Wang16}. Thus, conducting the experiment on a
  \texttt{x86\_64}-based Qemu virtual machine must be acknowledged as a
  limitation. It could even be that some of the undefined behaviors detected are
  not about the software experimented with it but about Qemu's (in)ability to
  emulate;~cf.~\cite{Jiang21}. Some bugs are also known to only occur on
  real~hardware~\cite{Choe25}. That said, it is difficult to say anything
  specific about the severity of this limitation except that it should be
  relatively minor because Qemu is supposedly particularly good at emulating
  \texttt{x86\_64}.}
\end{enumerate}

\section{Closing Remarks}\label{sec: closing remarks}

The results presented reiterate an argument from 2017 in which the authors
stated ``the obvious -- that, despite the many excellent advances in tooling
over the last few years'', problems with undefined behavior ``are far from
solved''~\cite{Cuoq17}.

Then, when having to live with undefined behavior, a recommendation to ``be very
careful, use good tools, and hope for the best''~\cite{Regehr10} seems
sensible. Even though the tooling has improved thanks to UBSan and other
sanitizers, using software usually rests on an implicit or explicit assumption
of others having been careful. Therefore, on the one hand, trusting others and
hoping for the best is what most desktop users can really only count on. For
developers, on the other hand, debugging and fixing undefined behavior may be
challenging~\cite{Xu23}, and, therefore, another recommendation~\cite{Wang16} to
turn on various related compiler options seems still sensible. When developing
new software with the C and C++ programming languages, it is a good idea to turn
also UBSan on by default.

\balance
\bibliographystyle{IEEEtran}

\begin{thebibliography}{10}
\providecommand{\url}[1]{#1}
\csname url@samestyle\endcsname
\providecommand{\newblock}{\relax}
\providecommand{\bibinfo}[2]{#2}
\providecommand{\BIBentrySTDinterwordspacing}{\spaceskip=0pt\relax}
\providecommand{\BIBentryALTinterwordstretchfactor}{4}
\providecommand{\BIBentryALTinterwordspacing}{\spaceskip=\fontdimen2\font plus
\BIBentryALTinterwordstretchfactor\fontdimen3\font minus
  \fontdimen4\font\relax}
\providecommand{\BIBforeignlanguage}[2]{{%
\expandafter\ifx\csname l@#1\endcsname\relax
\typeout{** WARNING: IEEEtran.bst: No hyphenation pattern has been}%
\typeout{** loaded for the language `#1'. Using the pattern for}%
\typeout{** the default language instead.}%
\else
\language=\csname l@#1\endcsname
\fi
#2}}
\providecommand{\BIBdecl}{\relax}
\BIBdecl

\bibitem{Ruohonen19RSDA}
J.~Ruohonen and K.~Rindell, ``{E}mpirical {N}otes on the {I}nteraction
  {B}etween {C}ontinuous {K}ernel {F}uzzing and {D}evelopment,'' in
  \emph{Proceedings of the IEEE International Symposium on Software Reliability
  Engineering Workshops (ISSREW 2019)}.\hskip 1em plus 0.5em minus 0.4em\relax
  Berlin: IEEE, 2019, pp. 276--281.

\bibitem{Song19}
D.~Song, J.~Lettner, P.~Rajasekaran, Y.~Na, S.~Volckaert, P.~Larsen, and
  M.~Fran, ``{S}o{K}: {S}anitizing for {S}ecurity,'' in \emph{Proceedings of
  the IEEE Symposium on Security and Privacy (S\&P)}.\hskip 1em plus 0.5em
  minus 0.4em\relax San Francisco: IEEE, 2019, pp. 1275--1295.

\bibitem{Oracle21}
E.~Zannoni, ``{I}mproving {A}pplication {S}ecurity with
  {U}ndefined{B}ehavior{S}anitizer {(UBSan)} and {GCC},'' 2021, {O}racle
  {L}inux {B}log, available online in 10 May 2026:
  \url{https://blogs.oracle.com/linux/improving-application-security-with-undefinedbehaviorsanitizer-ubsan-and-gcc}.

\bibitem{Packard25}
K.~Packard, ``{F}un {W}ith -fsanitize=undefined and {P}icolibc,'' 2025,
  {A}vailable online in 10 April 2026:
  \url{https://keithp.com/blogs/sanitizer-fun/}.

\bibitem{Wang12}
X.~Wang, H.~Chen, A.~Cheung, Z.~Jia, N.~Zeldovich, and M.~F. Kaashoek,
  ``{U}ndefined {B}ehavior: {W}hat {H}appened to my {C}ode?'' in
  \emph{Proceedings of the Asia-Pacific Workshop on Systems (APSYS
  2012)}.\hskip 1em plus 0.5em minus 0.4em\relax Seoul: ACM, 2021, pp. 1--7.

\bibitem{Clang26}
{The Clang Team}, ``{U}ndefined{B}ehavior{S}anitizer,'' 2026, {A}vailable
  online in 10 April 2026:
  \url{https://clang.llvm.org/docs/UndefinedBehaviorSanitizer.html}.

\bibitem{GCC26a}
{Free Software Foundation, Inc.}, ``3.13 {P}rogram {I}nstrumentation
  {O}ptions,'' 2026, {A}vailable online in 10 April 2026:
  \url{https://gcc.gnu.org/onlinedocs/gcc/Instrumentation-Options.html}.

\bibitem{WG14}
{WG14/N1256}, ``{C}ommittee {D}raft -- {S}eptermber 7, 2007 {ISO}/{IEC}
  9899:{TC3},'' 2007, {I}nternational standardization working group for the
  programming language C. Available online in 11 April 2026:
  \url{https://www.open-std.org/jtc1/sc22/WG14/www/docs/n1256.pdf}.

\bibitem{WG12}
T.~Doumler and J.~Berne, ``{A} {F}ramework for {S}ystematically {A}ddressing
  {U}ndefined {B}ehaviour in the {C++} {S}tandard,'' 2025, {JTC1}/{SC22}/{WG21}
  -- {T}he {C++} {S}tandards {C}ommittee -- {ISOCPP}, available online in May
  2026:
  \url{https://www.open-std.org/jtc1/sc22/wg21/docs/papers/2025/p3100r5.pdf}.

\bibitem{Wang16}
X.~Wang, N.~Zeldovich, M.~F. Kaashoek, and A.~Solar-Lezama, ``{A}
  {D}ifferential {A}pproach to {U}ndefined {B}ehavior {D}etection,''
  \emph{Communications of the ACM}, vol.~59, no.~3, pp. 99--106, 2016.

\bibitem{Hathorn15}
C.~Hathhorn, C.~Ellison, and G.~Ros, ``{D}efining the {U}ndefinedness of {C},''
  in \emph{Proceedings of the 36th ACM SIGPLAN Conference on Programming
  Language Design and Implementation (PLDI 2015)}.\hskip 1em plus 0.5em minus
  0.4em\relax Portland: ACM, 2015, pp. 336--345.

\bibitem{Xu23}
J.~Xu, K.~Lu, Z.~Du, Z.~Ding, L.~Li, Q.~Wu, M.~Payer, and B.~Mao, ``{S}ilent
  {B}ugs {M}atter: {A} {S}tudy of {C}ompiler-{I}ntroduced {S}ecurity {B}ugs,''
  in \emph{Proceedings of the 32nd USENIX Security Symposium (USENIX
  2023)}.\hskip 1em plus 0.5em minus 0.4em\relax Anaheim: USENIX, 2023, pp.
  3655--3672.

\bibitem{Popescu25}
L.~Popescu and N.~P. Lopes, ``{E}xploiting {U}ndefined {B}ehavior in {C/C++}
  {P}rograms for {O}ptimization: {A} {S}tudy on the {P}erformance {I}mpact,''
  \emph{Proceedings of the ACM on Programming Languages}, vol.~9, no. PLDI, pp.
  348--371, 2025.

\bibitem{Winters20}
T.~Winters, T.~Manshreck, and H.~Wright, \emph{{S}oftware {E}ngineering at
  {G}oogle: {L}essons {L}earned {F}rom {P}rogramming {O}ver {T}ime}.\hskip 1em
  plus 0.5em minus 0.4em\relax Sebastopol: O'Reilly, 2020.

\bibitem{Nong20}
Y.~Nong and H.~Cai, ``{A} {P}reliminary {S}tudy on {O}pen-{S}ource {M}emory
  {V}ulnerability {D}etectors,'' in \emph{Proceedings of the IEEE 27th
  International Conference on Software Analysis, Evolution and Reengineering
  (SANER 2020)}, London, 2020, pp. 557--561.

\bibitem{Vyukovaa20}
N.~I. V\'yukovaa, V.~A. Galatenkoa, and S.~V. Samborskii, ``{D}ynamic {P}rogram
  {A}nalysis {T}ools in {GCC} and {CLANG} {C}ompilers,'' \emph{Programming and
  Computer Software}, vol.~46, pp. 81--296, 2020.

\bibitem{Zhang21}
J.~Zhang, S.~Wang, M.~Rigger, P.~He, and Z.~Su, ``{S}an{R}azo{R}: {R}educing
  {R}edundant {S}anitizer {C}hecks in {C}/{C}++ {P}rograms,'' in
  \emph{Proceedings of the 15th USENIX Symposium on Operating Systems Design
  and Implementation}.\hskip 1em plus 0.5em minus 0.4em\relax Online: USENIX,
  2021, pp. 479--494.

\bibitem{Li24}
S.~Li and Z.~Su, ``{UBF}uzz: {F}inding {B}ugs in {S}anitizer
  {I}mplementations,'' pp. 435--449, 2024.

\bibitem{Rocca25}
E.~Rocca, ``{P}ackages {N}ot {U}sing {T}he {D}efault {B}uild {F}lags: {A}
  {T}axonomy,'' in \emph{Proceedings of the 26th Debian Conference (DebConf
  2025)}, Brest, 2025, pp. 1--5, available online in 11 April 2026:
  \url{https://hal.science/hal-05334704/document}.

\bibitem{Ruohonen25PST}
J.~Ruohonen, M.~Saddiqa, and K.~Sierszecki, ``{A} {S}tatic {A}nalysis of
  {P}opular {C} {P}ackages in {L}inux,'' in \emph{Proceedings of the 22nd
  Annual International Conference on Privacy, Security, and Trust (PST
  2025)}.\hskip 1em plus 0.5em minus 0.4em\relax Fredericton: IEEE, 2025, pp.
  1--10.

\bibitem{Wohlin24}
C.~Wohlin, P.~Runeson, M.~H\"ost, M.~C. Ohlsson, B.~Regnell, and A.~Wessl\'en,
  \emph{{E}xperimentation in {S}oftware {E}ngineering}, 2nd~ed.\hskip 1em plus
  0.5em minus 0.4em\relax Heidelberg: Springer, 2024.

\bibitem{Gentoo26a}
{Gentoo}, ``{M}inimal {I}nstallation {CD},'' 2026, available online in 10 April
  2026:
  \url{https://distfiles.gentoo.org/releases/amd64/autobuilds/20260408T183104Z/install-amd64-minimal-20260408T183104Z.iso}.

\bibitem{Gentoo26b}
------, ``{D}efault {S}tage {A}rchives: {S}tage {D}esktop {P}rofile \&
  {O}pen{RC},'' 2026, available online in 10 April 2026:
  \url{https://distfiles.gentoo.org/releases/amd64/autobuilds/20260410T130145Z/stage3-amd64-desktop-openrc-20260410T130145Z.tar.xz}.

\bibitem{Gentoo26c}
------, ``{G}entoo {AMD64} {H}andbook,'' 2026, available online in 10 April
  2026: \url{https://wiki.gentoo.org/wiki/Handbook:AMD64}.

\bibitem{Gentoo26d}
------, ``{U}ndefined{B}ehavior{S}anitizer,'' 2026, available online in 10
  April 2026: \url{https://wiki.gentoo.org/wiki/UndefinedBehaviorSanitizer}.

\bibitem{GCC26b}
{Free Software Foundation, Inc.}, ``{C} {S}tandards {S}upport in {GCC},'' 2026,
  {A}vailable online in 10 June 2026:
  \url{https://gcc.gnu.org/projects/c-status.html#c23}.

\bibitem{GCC26c}
------, ``{C}++ {S}tandards {S}upport in {GCC},'' 2026, {A}vailable online in
  10 June 2026: \url{https://gcc.gnu.org/projects/cxx-status.html?#cxx17}.

\bibitem{Ruohonen25ICTSSa}
J.~Ruohonen, ``{A} {T}ime {S}eries {A}nalysis of {A}ssertions in the {L}inux
  {K}ernel,'' in \emph{Proceedings of the 37th International Conference on
  Testing Software and Systems (ICTSS 2025)}.\hskip 1em plus 0.5em minus
  0.4em\relax Limassol: Springer, 2026, pp. 3--15.

\bibitem{RedHat21}
J.~Kratochvil, ``{M}emory {E}rror {C}hecking in {C} and {C}++: {C}omparing
  {S}anitizers and {V}algrind,'' 2021, {R}ed {H}at {D}eveloper {B}log,
  available online in 10 April 2026:
  \url{https://developers.redhat.com/blog/2021/05/05/memory-error-checking-in-c-and-c-comparing-sanitizers-and-valgrind}.

\bibitem{Ayala22}
C.~Ayala, B.~Turhan, X.~Franch, and N.~Juristo, ``{U}se and {M}isuse of the
  {T}erm ``{E}xperiment'' in {M}ining {S}oftware {R}epositories {R}esearch,''
  \emph{IEEE Transactions on Software Engineering}, vol.~48, no.~11, pp.
  4229--4248, 2022.

\bibitem{Haas21}
R.~Haas, D.~Elsner, E.~Juergens, A.~Pretschner, and S.~Apel, ``{H}ow {C}an
  {M}anual {T}esting {P}rocesses {B}e {O}ptimized? {D}eveloper {S}urvey,
  {O}ptimization {G}uidelines, and {C}ase {S}tudies,'' in \emph{Proceedings of
  the 29th ACM Joint Meeting on European Software Engineering Conference and
  Symposium on the Foundations of Software Engineering (ESEC/FSE 2021)}.\hskip
  1em plus 0.5em minus 0.4em\relax ACM, 2021, pp. 1281--1291.

\bibitem{Hemmati17}
H.~Hemmati, Z.~Fang, M.~V. M\"antylä, and B.~Adams, ``{P}rioritizing {M}anual
  {T}est {C}ases in {R}apid {R}elease {E}nvironments,'' \emph{Journal of
  Software: Testing, Verification and Reliability}, vol.~27, no.~6, p. e1609,
  2017.

\bibitem{Itkonen07}
J.~Itkonen, M.~V. M\"antyla, and C.~Lassenius, ``{D}efect {D}etection
  {E}fficiency: {T}est {C}ase {B}ased vs.~{E}xploratory {T}esting,'' in
  \emph{Proceedings of he First International Symposium on Empirical Software
  Engineering and Measurement (ESEM 2007)}.\hskip 1em plus 0.5em minus
  0.4em\relax Madrid: IEEE, 2007, pp. 61--70.

\bibitem{Zhang09}
H.~Zhang, ``{A}n {I}nvestigation of the {R}elationships {B}etween {L}ines of
  {C}ode and {D}efects,'' in \emph{Proceedings of the IEEE International
  Conference on Software Maintenance (ICSM 2009)}, Edmonton, 2009, pp.
  274--283.

\bibitem{Kampenes09}
V.~B. Kampenes, T.~Dyb\r{a}, J.~E. Hannay, and D.~I.~K. Sj\o{}berg, ``{A}
  {S}ystematic {R}eview of {Q}uasi-{E}xperiments in {S}oftware {E}ngineering,''
  \emph{Information and Software Technology}, vol.~51, pp. 71--82, 2009.

\bibitem{Jiang21}
M.~Jiang, T.~Xu, Y.~Zhou, Y.~Hu, M.~Zhong, L.~Wu, X.~Luo, and K.~Ren,
  ``{A}utomatically {L}ocating {ARM} {I}nstructions {D}eviation {B}etween
  {R}eal {D}evices and {CPU} {E}mulators,'' 2021, archived manuscript,
  available online in 12 April 2026: \url{https://arxiv.org/abs/2105.14273}.

\bibitem{Choe25}
W.~Choe, R.~Wang, A.~Benazir, and F.~X. Lin, ``{P}roto: {A} {G}uided {J}ourney
  {T}hrough {M}odern {OS} {C}onstruction,'' in \emph{Proceedings of the ACM
  SIGOPS 31st Symposium on Operating Systems Principles (SOSP 2025)}.\hskip 1em
  plus 0.5em minus 0.4em\relax Seoul: ACM, 2025, pp. 50--66.

\bibitem{Cuoq17}
P.~Cuoq and J.~Regehr, ``{U}ndefined {B}ehavior in 2017,'' 2017, available
  online in June 2026: \url{https://blog.regehr.org/archives/1520}.

\bibitem{Regehr10}
J.~Regehr, ``{A} {G}uide to {U}ndefined {B}ehavior in {C} and {C}++, {P}art
  1,'' 2010, available online on 12 April 2026:
  \url{https://blog.regehr.org/archives/213}.

\end{thebibliography}


\end{document}